# Where Do Thin Tails Come From?

## (Studies in (ANTI)FRAGILITY)

Nassim N. Taleb

The literature of heavy tails starts with a random walk and finds mechanisms that lead to fat tails under aggregation. We follow the inverse route and show how starting with fat tails we get to thin-tails when deriving the probability distribution of the response to a random variable. We introduce a general dose-response curve and argue that the left and right-boundedness (or saturation) of the response in natural settings leads to thin-tails, even when the "underlying" random variable at the source of the exposure is fat-tailed.

Very Preliminary Version, July 2013.

## The Origin of Thin Tails.

We have imprisoned the "statistical generator" of things on our planet into the random walk theory: the sum of i.i.d. variables eventually leads to a Gaussian, which is an appealing theory. Or, actually, even worse: at the origin lies a simpler Bernoulli binary generator with variations limited to the set {0,1}, normalized and scaled, under summation. Bernoulli, De Moivre, Galton, Bachelier: all used the mechanism, as illustrated by the Quincunx in which the binomial leads to the Gaussian, either for pedagogy or conviction. This has traditionally been the "generator" mechanism behind most processes, from Brownian motion to martingales. About every standard textbook hints at the "naturalness" of the thus-obtained Gaussian, or take it for granted.

In that sense, powerlaws are pathologies. Traditionally, the tendency for researchers has been to "justify" fat tailed distributions using the canonical random walk generator, but twinging it thanks to a series of mechanisms that start with an aggregation of random variables that does not lead to the central limit theorem, owing to lack of independence and the magnification of moves through some mechanism of contagion: preferential attachment, comparative advantage, and similar mechanisms[1] . (Few research traditions, such as the works in complex systems, escape it.)

But the random walk theory fails to accommodate some obvious phenomena.

    First, many things move by jumps and discontinuities that cannot come from the random walk and the conventional Brownian motion, a theory that proved to be sticky[2] .

    Second, consider the distribution of the size of animals in nature, considered within-species. The height and weight of humans follow (almost) a Normal Distribution but it is hard to find mechanism of random walk behind it (this is an observation imparted to the author by Yaneer Bar-Yam).

    Third, uncertainty and opacity lead to power laws, when a statistical mechanism has an error rate which in turn has an error rate, and thus, recursively[3] .

Our approach here is to assume that the "source" random variables, under absence of constraints, are power law-distributed. This is the default in the absence of boundedness or compactness. Then, the *response*, that is, a function of the source random variable, considered in turn as an "inherited" random variable, will have its own properties. If the response is bounded, then the dampening of the tails of the inherited distribution will lead it to bear the properties of the Gaussian, or the class of distributions possessing finite moments of all orders.

## The Dose Response

Let us start with case of the bounded sigmoid function, and generalize to cover broad cases. By "dose-response" we cover stress or other inputs as part of the dose.

Let $S^N(x): \mathbb{R} \to [0 \wedge K_L, K_R]$ be a continuous function possessing derivatives $(S^N)^{(n)}(x)$ of all orders, expressed as an $N$-summed and scaled standard sigmoid functions:

$$S^N(x) \equiv \sum_{k=1}^{N} \frac{a_k}{1 + \exp(-b_k x + c_k)} + K_L \tag{1}$$

where $a_k$, $b_k$, $c_k$ are norming constants $\in \mathbb{R}$, satisfying:

i) $S^N(-\infty) = K_L$

ii) $S^N(\infty) = K_R$

where $K_R = \sum_{i=1}^{N} a_k + K_L$

and (equivalently for the first and last of the following conditions)

iii) $\frac{\partial^2 S^N}{\partial x^2} \geq 0$ for $x \in (-\infty, x_1)$, $\frac{\partial^2 S^N}{\partial x^2} < 0$ for $x \in (x_2, x_{>2})$, and $\frac{\partial^2 S^N}{\partial x^2} \geq 0$ for $x \in (x_{>2}, \infty)$, with $x_1 > x_2 \geq x_3 \ldots \geq x_N$.

Assume $K_L = 0$. The shapes at different calibrations are shown in Figure 1, in which we combined different values of N=2 $S^2(x, a_1, a_2, b_1, b_2, c_1, c_2)$, and the standard sigmoid $S^1(x, a_1, b_1, c_1)$, with $a_1=1$, $b_1=1$ and $c_1=0$. As we can see, unlike the common sigmoid, the asymptotic response can be lower than the maximum, as our curves are not monotonically increasing. The sigmoid shows benefits increasing rapidly (the convex phase), then increasing at a slower and slower rate until saturation. Our more general case starts by increasing, but the response can be actually negative beyond the saturation phase, though in a convex manner. Harm slows down and becomes "flat" when something is totally broken.

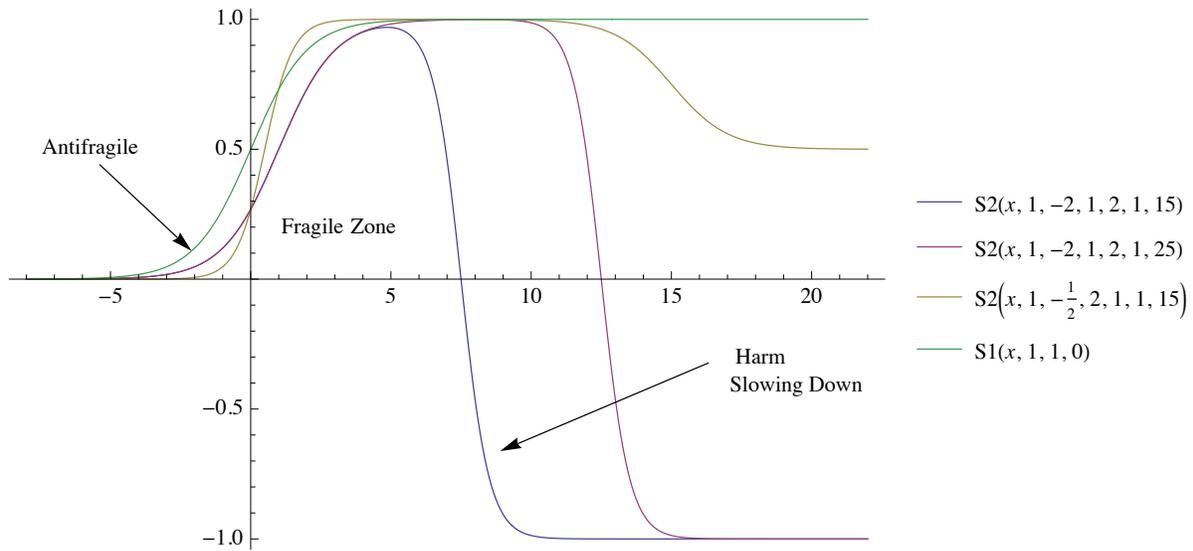

**Figure 1.** The Generalized Response Curve, special cases: $S^2(x, a_1, a_2, b_1, b_2, c_1, c_2)$, $S^1(x, a_1, b_1, c_1)$. The convex part with positive first derivative has been designated as "antifragile"

Parameter $a$ sets the height, or $K_R - (0 \wedge K_L)$, $b$ sets the variance, or the slope of the sigmoid, and $c$ sets the displacement

Note that the convex part of the graph corresponds to the "antifragile" exposure, i.e. gains from local stochasticity, disturbances and variance around a given mean (by Jensen's Inequality), while the "fragile" case is harmed by it.

The same framework but with opposite characteristics than the sigmoid, namely the probit or inverse cumulative Gaussian, can model fat-tailedness as a convex positive response and a concave negative one, in situations usually mapped as cumulative advantage or preferential attachment.

## Properties of the Inherited Probability Distribution

Now let $x$ be a random variable distributed according to a general fat tailed distribution, with power laws at large negative and positive values, expressed (for clarity, without loss of generality) as a Student T Distribution with scale $\sigma$ and exponent $\alpha$, and support on the real line. Its domain $\mathcal{D}^f = (-\infty, \infty)$, and density $f_{\sigma,\alpha}(x)$:

$$f_{\sigma,\alpha}(x) \equiv \frac{\left(\frac{\alpha}{\alpha + \frac{x^2}{\sigma^2}}\right)^{\frac{1+\alpha}{2}}}{\sqrt{\alpha}\ \sigma\ B\left[\frac{\alpha}{2}, \frac{1}{2}\right]} \tag{2}$$

where $B$ is the Euler Beta function, $B(a, b) = \Gamma(a)\Gamma(b)/\Gamma(a+b) = \int_0^1 t^{a-1}(1-t)^{b-1}\,dt$.

An illustrative simulation of the convex-concave transformations of the terminal probability distribution is shown in Figure 2, with four cases considered.

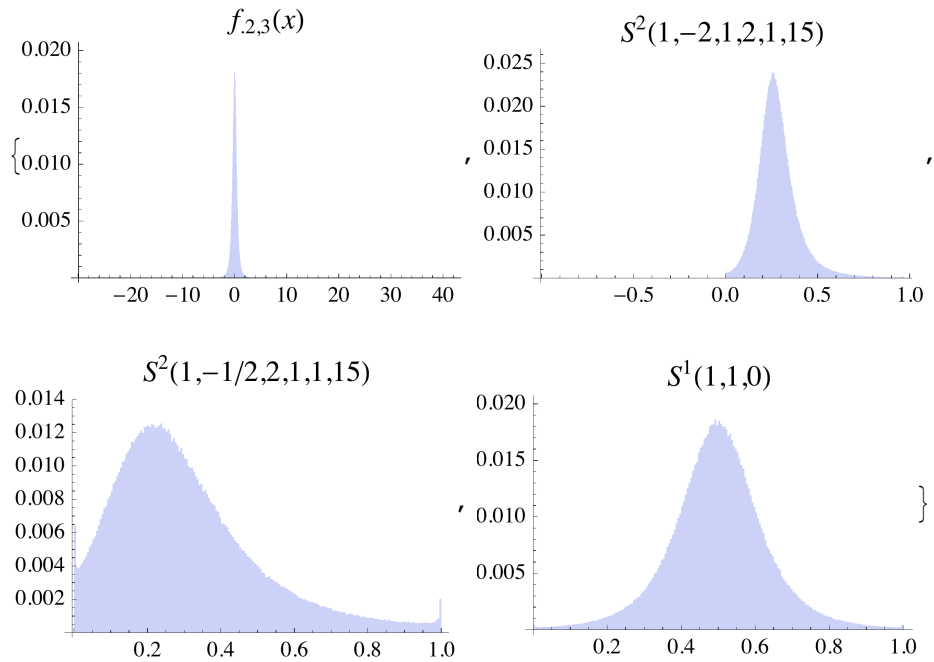

**Figure 2.** Histogram for the different inherited probability distributions (simulations, $N = 10^6$)

We can see that the Kurtosis of the inherited distributions drops at higher $\sigma$ thanks to the boundedness of the payoff, making the truncation to the left and the right meaningful, as a Dirac-Delta mass forms at the points $K_L$ and $K_R$. Kurtosis for $f_{.2,3}$ is infinite, but in-sample will be extremely high, but, of course, finite. So we use it as a benchmark to see for a given sample the drop from the calibration of the response curves.

| Distribution | Kurtosis |
|---|---|
| $f_{.2,3}(x)$ | 86.3988 |
| $S^2(1,-2,1,2,1,15)$ | 8.77458 |
| $S^2(1,-1/2,2,1,1,15)$ | 4.08643 |
| $S^1(1,1,0)$ | 4.20523 |

**Analytical Derivation:** We start with the case of the standard sigmoid, i.e., $N = 1$

$$S(x) \equiv \frac{a_1}{1+\exp(-b_1 x + c_1)}$$

g(x) is the inherited distribution, which can be shown to have a scaled domain $\mathcal{D}^g = ((0 \wedge K_L), K_R)$. It becomes:

$$g(x) = \frac{a1 \left( \frac{\alpha}{\alpha + \frac{\left(\log\left(\frac{x}{a1-x}\right)+c1\right)^2}{b1^2 \sigma^2}} \right)^{\frac{\alpha+1}{2}}}{\sqrt{\alpha}\; b1\, \sigma\, x\, B\left(\frac{\alpha}{2},\frac{1}{2}\right)(a1-x)} \tag{3}$$

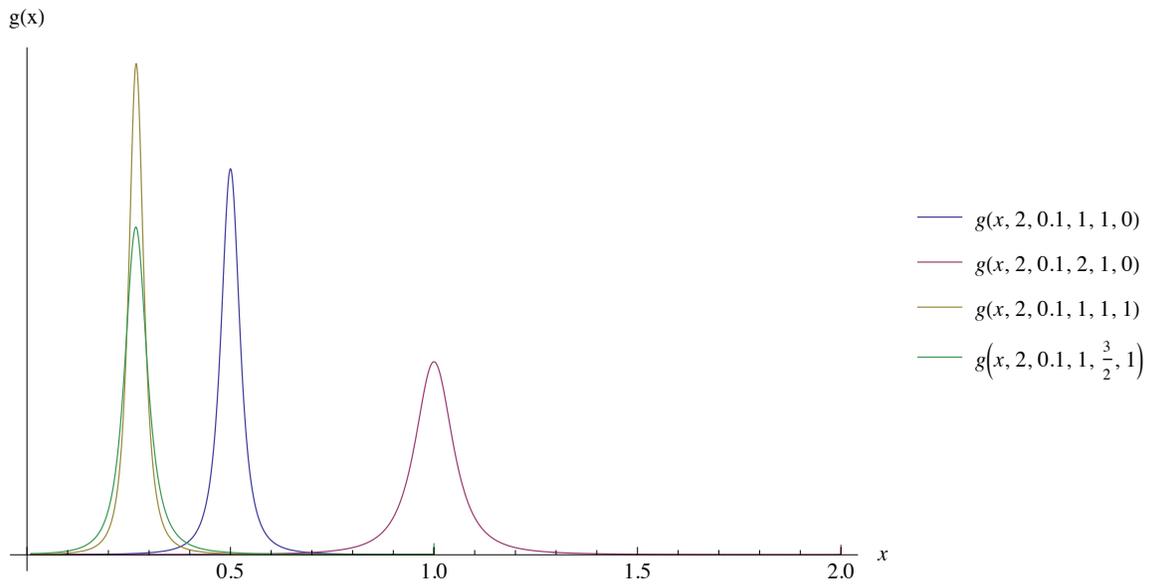

**Figure 3.** The different inherited probability distributions.

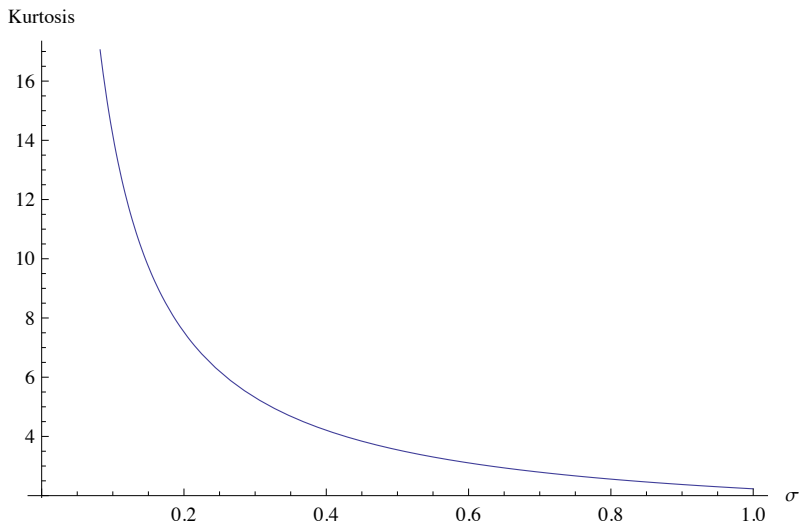

**Figure 4.** The Kurtosis of the standard drops along with the scale $\sigma$ of the power law

***Remark 1***: *The inherited distribution from S(x) will have a compact support regardless of the probability distribution of x.*

For higher values of $N$, the inverse function of $S^{>1}(x)$ is not analytic, thus disallowing explicit expressions of the probability distributions, but simulations show the properties to be not too different.

## Recovering the Gaussian (The Unbounded Case)

Now consider the more special case of recovering the Gaussian. Let us set $P_{s,m}(.)$ as the cumulative density function and $p_{s,m}(.)$ the density of a Gaussian with mean $m$ and standard deviation $s$, and $F_{\sigma,\alpha}(.)$ the cumulative for the symmetric power law distribution we saw earlier, with density $f_{\sigma,\alpha}(.)$. Both domains are $\mathbb{R}$. Let $y = \gamma(x)$ be a monotone increasing (and differentiable) function.

Setting $\int_{-\infty}^{y} p_{s,m}(z)\, dz = \int_{-\infty}^{x} f_{\sigma,\alpha}(z)\, dz$ yields the following solutions

For the square exponent, $\alpha=2$,

$$\gamma(x)\mid_{\alpha=2} = m - \sqrt{2}\, s\, \mathrm{erfc}^{-1}\left(\frac{\sigma x \sqrt{\frac{2\sigma^2 + x^2}{\sigma^2}}}{2\sigma^2 + x^2} + 1\right) \tag{4}$$

and for the cubic exponent $\alpha=3$,

$$\gamma(x)\mid_{\alpha=3} = m - \sqrt{2}\, s\, \mathrm{erfc}^{-1}\left(\frac{\frac{2\sqrt{3}\,\sigma x}{3\sigma^2 + x^2} + 2\tan^{-1}\left(\frac{x}{\sqrt{3}\,\sigma}\right) + \pi}{\pi}\right) \tag{5}$$

Where erfc is the complementary error function. Predictably both $\gamma(x)\mid_{\alpha=2}$ and $\gamma(x)\mid_{\alpha=3}$ are unbounded sigmoids.

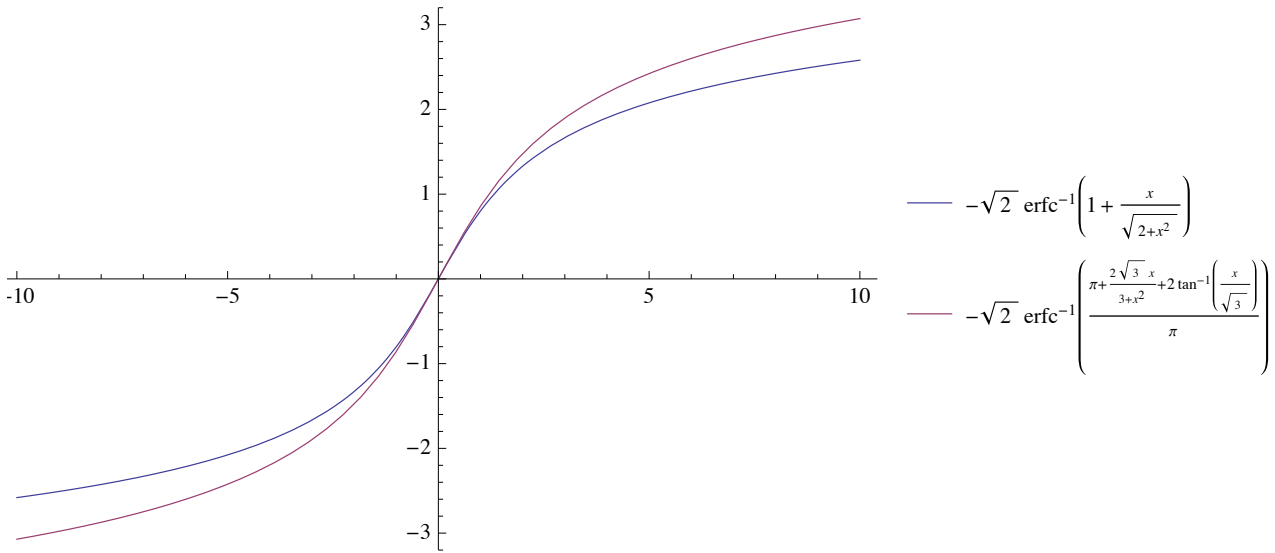

**Figure 5.** The different roads that lead to the Gaussian

## Conclusion and Remarks

We showed the possibility of the response (dose-response or stress-response) to variations in an ecology as the neglected origin of the thin-tailedness of observed distributions in nature. This approach to the dose-response curve is quite general, and can be used outside biology (say in the Kahneman-Tversky prospect theory, in which their version of the utility concept with respect to changes in wealth is convex on the left, because unhappiness is bounded by death, and concave on the right).

## Acknowledgments

Yaneer Bar-Yam, Jim Gatheral, Raphael Douady, Carl Fakhry, Brent Halonen.

...